\documentclass[10pt,amsmath,twocolumn,superscriptaddress,aps,prb,floatfix]{revtex4-1} %nofootinbib
\usepackage{amsthm,amsfonts,graphicx,verbatim, color}
\usepackage{bm}% bold math
\usepackage[utf8]{inputenc}
\usepackage{xcolor}
\usepackage{xifthen}
\usepackage{siunitx}
\usepackage[normalem]{ulem}
\usepackage[colorlinks, linkcolor=darkblue, citecolor=darkblue, urlcolor=darkblue]{hyperref}
\usepackage{soul}

\let\oldmarginpar\marginpar
\renewcommand\marginpar[1]{\-\oldmarginpar[\raggedleft\footnotesize #1]%
{\raggedright\footnotesize #1}}

\newcommand{\Ref}[1]{Ref.~\cite{#1}}

\newcommand{\be}{\begin{equation}}
\newcommand{\ee}{\end{equation}}
\newcommand{\bea}{\begin{equation}}
\newcommand{\eea}{\end{equation}}

\newcommand{\Tr}{\mathrm{ Tr}\,}
\renewcommand{\epsilon}{\varepsilon}

\renewcommand{\Im}{\mathrm{ Im}\,}

\newcommand{\barptt}[1]{\bar{\mathrm{#1}}}
\newcommand{\barpt}[1]{$\barptt{#1}$}

\newcommand{\bs}[1]{\boldsymbol{#1}}

\definecolor{darkblue}{rgb}{0,0,.5}
\definecolor{orange}{rgb}{1,.4,0}
\newcommand{\red}[1]{{\textcolor{red}{#1}}}

\newcommand{\magenta}[1]{{\textcolor{magenta}{#1}}}

\newcommand{\new}[2][]{\ifthenelse{\isempty{#1}}{}{\textcolor{orange}{\sout{#1}}\ifthenelse{\isempty{#2}}{}{ }}\textcolor{darkblue}{#2}}
\newcommand{\todo}[2][]{\magenta{TODO\ifthenelse{\isempty{#1}}{}{ (#1)}:} \red{#2}}

\newcommand{\affuzh}{\affiliation{Department of Physics, University of Zurich, Winterthurerstrasse 190, 8057 Zurich, Switzerland}}

\begin{document}
\title{Magnetic Weyl and Dirac Kondo semimetal phases in heterostructures}
\author{Seulgi Ok}\affuzh
\author{Markus Legner}
\affiliation{Institute for Theoretical Physics, ETH Zurich, 8093 Zurich, Switzerland}
\author{Titus Neupert}\affuzh
\author{Ashley M.\ Cook}\affuzh
\begin{abstract}
We study a layered three-dimensional heterostructure in which two types of Kondo insulators are stacked alternatingly. 
One of them is the topological Kondo insulator SmB\textsubscript{6}, the other one an isostructural Kondo insulator $A$B\textsubscript{6}, where $A$ is a rare-earth element, e.g., Eu, Yb, or Ce. We find that if the latter orders ferromagnetically, the heterostructure generically becomes a magnetic Weyl Kondo semimetal, while antiferromagnetic order can yield a magnetic Dirac Kondo semimetal. We detail both scenarios with general symmetry considerations as well as concrete tight-binding calculations and show that type-I as well as type-II magnetic Weyl/Dirac Kondo semimetal phases are possible in these heterostructures. Our results demonstrate that Kondo insulator heterostructures are a versatile platform for design of strongly correlated topological semimetals.
 
\end{abstract}
\maketitle

\section{Introduction}
Recent efforts to study phases defined by topologically-protected band structure degeneracies~\cite{weng2015,burkov2011,huang2015,xu2015,lv2015, wang2012,liu2014,borisenko2014,yang2014,chang2016,xu2015',young2016, wang2017, wang2017two} reflect their great potential for technological applications~\cite{pikulin2016,rajamathi2016} and for table-top experiments on quasiparticles both analogous to high-energy particles and beyond this framework ~\cite{bradlyn2016,kane2016, zhu2016, bzdusek2016,chiu2014,xie2015,yu2015,bian2015, wieder2016}. There is also a major undertaking to study these systems in the presence of correlations that suggests far richer physics beyond the weakly-correlated regime~\cite{lu2015, liu2016, nomura2014, lundgren2015, gonzalez2015, braguta2016, zhai2016, roy20162, wurff2016, roy2016,wei2012,schaffer2015, hermanns2015}.

Heavy-fermion systems serve as an important guide in the study of correlations, and therefore of correlated topological phases: 
%\blue{ref1: Authors use non-interacting tight-binding models and consider the interaction of the topological surface states through a trivial insulator. This is mainly the proposal of Burkov and Balents (Ref: 2). They also consider various magnetic insulators as buffer layers to study various other semimetallic phases. Even though this goes beyond Ref: 2, it is not such a new development. There are already many papers that considers such a scenario in the context of non-interacting systems. ref1 comment2: Authors emphasise that Kondo insulators are strongly correlated insulators. However their analysis is limited to non-interacting tight binding models. There is no connection with the premise of the paper with the actual calculations.} 
There is evidence these compounds host topological Kondo phases, which share the topological classification of weakly-correlated topological phases but with the essential difference that it is the strongly correlated Kondo effect that protects the topologically non-trivial state.~\cite{chen2016}. This area of research began with topological Kondo insulators~\cite{dzero2010,alexandrov2013,iaconis2015,feng2013, li2014,min2014, miyazaki2012, wolgast2013, zhang2013,xu2013, neupane2013, frantzeskakis2013, takimoto2011, tran2012,lu2013, dzero2013, deng2013} but has more recently been extended to Chern Kondo insulators~\cite{chen2016},  topological crystalline Kondo insulators~\cite{ye2013,weng2014}, M\"obius Kondo insulators\cite{chang2017}, Weyl Kondo semimetals due to inversion symmetry-breaking~\cite{lai2016} and non-magnetic Dirac Kondo semimetals~\cite{feng2016}. 

%\blue{(ref2: \emph{Weyl Kondo semimetal has recently been proposed in bulk} (3D) Kondo lattice systems that break inversion symmetry. It seems that the construction here, invoking layered heterostructures, lead to a Weyl Kondo semimetal state that is similar in nature.}

The characterization of topological Kondo semimetals remains incomplete, however. The magnetic Weyl semimetal---a Weyl semimetal realized via time-reversal symmetry-breaking as opposed to inversion symmetry-breaking\cite{burkov2011}---still lacks a counterpart phase, the magnetic Weyl Kondo semimetal, which is characterized by Weyl cones in the bulk electronic structure that are protected by the Kondo effect. %This is the essential difference between a topological Kondo phase and a weakly-correlated topological phase~\cite{chen2016}.

The magnetic Weyl semimetal---not the Weyl semimetal due to inversion symmetry-breaking---can be the elementary example of a Weyl semimetal in the sense that it may realize the minimum number of two Weyl cones~\cite{burkov2011}. 
Experimental signatures of magnetic Weyl semimetals are simpler and may differ from those of Weyl semimetals due to inversion symmetry-breaking~\cite{cano2017}. 
To date, there has only been one realization of the magnetic Weyl semimetal, and only in the presence of an applied magnetic field, while there have been many experimental studies of Weyl semimetals due to inversion symmetry-breaking~\cite{hirschberger2016}.

The magnetic Dirac Kondo semimetal is a state possessing Dirac cones, i.e., four-fold band degeneracies with linear dispersion in all directions in momentum space, in the bulk electronic structure when time-reversal symmetry is broken. These degeneracies emerge as a result of the Kondo effect and are protected by crystal symmetries of the magnetic space group. Magnetic Dirac semimetals have only been discussed very recently~\cite{wieder2016,tang2016, wang2017two, wang2017}. 
%These phases are opportunities to study the interplay between Dirac physics and magnetism, a magnetic Dirac semimetal existing as a result of strong correlations, such as the magnetic Dirac Kondo semimetal.

\begin{figure}[t]
\centering
\includegraphics[width=0.46\textwidth]{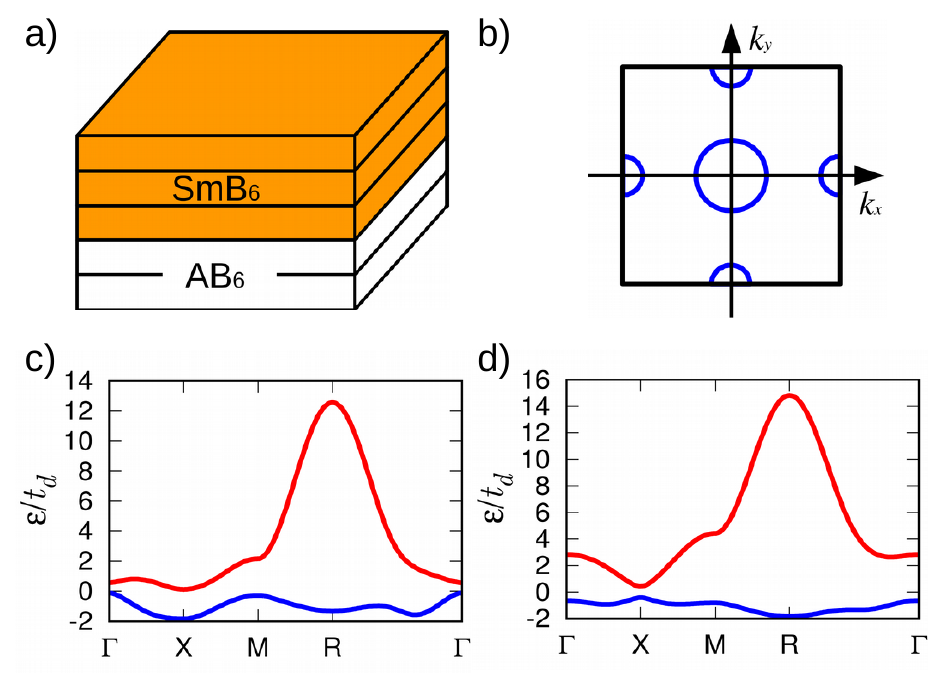}
\caption{
a) Unit cell of the layered heterostructure consisting of a finite number of layers of each of two isostructural Kondo insulators, here the topological Kondo insulator SmB\textsubscript{6} and the trivial narrow-gap insulator $A$B\textsubscript{6}. 
b) Schematic of the Fermi surface for some chemical potential $\mu$ intersecting the three Dirac cones on the (001) surface of SmB\textsubscript{6}.
Bulk electronic structures of the tight-binding models~\eqref{eq: lattice hamiltonian} for c) SmB\textsubscript{6} and d) an example $A$B\textsubscript{6} compound, respectively, arising from the hybridization between flat $f$ bands and dispersive $d$ bands, although the electronic structure of $A$B\textsubscript{6} could more generally also include $p$-bands near the Fermi level or just $p$- and $d$-bands near the Fermi level, or even be metallic~\cite{yan2017}, so long as the requisite symmetries and parent compound topologies are present, although this latter case is not ideal.
}
\label{fig1}
\end{figure}

In this work, we show that both the magnetic Weyl Kondo semimetal and magnetic Dirac Kondo semimetal may, conveniently, be constructed from the topologically non-trivial surface states of the topological Kondo insulator samarium hexaboride (SmB\textsubscript{6}), the most well-studied Kondo insulator and one of the best topological insulators when benchmarked by the minimal bulk conductivity~\cite{rakoski2017}. In this work, we explore the potential that SmB\textsubscript{6} and a family of isostructural materials $A$B\textsubscript{6} (where $A$ is a rare-earth element or a combination of two dopants) hold for designing topological metals when stacked in a periodically repeating heterostructure [see Fig.~\ref{fig1} a)]. 
%\blue{ref1: Authors use non-interacting tight-binding models and consider the interaction of the topological surface states through a trivial insulator. This is mainly the proposal of Burkov and Balents (Ref: 2). They also consider various magnetic insulators as buffer layers to study various other semimetallic phases. Even though this goes beyond Ref: 2, it is not such a new development. There are already many papers that considers such a scenario in the context of non-interacting systems.  ref1 comment2: Authors emphasise that Kondo insulators are strongly correlated insulators. However their analysis is limited to non-interacting tight binding models. There is no connection with the premise of the paper with the actual calculations.} 
As these topological phases are constructed directly from the surface states of a topological Kondo insulator and these surface states are protected by the Kondo effect, these phases are topological \textit{Kondo} metals.

We focus on symmetry-protected Weyl and Dirac cones brought about by ferromagnetic and antiferromagnetic order, respectively. While based on a proposal for weakly-correlated Weyl semimetals~\cite{burkov2011}, our work goes beyond this proposal by 
%\blue{(ref2: Kondo/Anderson lattice models give rise to strong renormalizations as a result of the large Coulomb repulsion. Such effect does not seem to have been analyzed in the paper.} 
(i) instead deriving topological semimetal phases from topologically non-trivial electronic structures resulting from intrinsically strongly-correlated Kondo physics~\cite{chen2016}, 
(ii) including the case of three dimensional magnetic Dirac Kondo semimetals, 
(iii) including realization of both type-I and type-II Weyl/Dirac Kondo semimetal phases, and 
(iv) utilizing the strong correlations of hexaborides as the sources of requisite magnetic orders as opposed to magnetic dopants~\cite{Barla2005, Derr2008}. We note that SmB$_6$ furthermore exhibits evidence of exotic correlated topological surface states~\cite{Nakajima2016, thomson2016, alexandrov2015, roy2015, nikolic2014}, which could enrich the physics of topological Kondo semimetals constructed from SmB$_6$ surface states.

%\blue{ref1: They even claim that EuB6 is a ferromagnetic insulator. It is very well known that EuB6 is a ferro-magnetic metal. I strongly suggest the authors to look at the transport properties of the materials before they use them in their manuscripts.} \blue{ref1: Authors argue that antiferromagnetism is a possible magnetic order since SmB6 becomes magnetic under pressure. It is true that SmB6 becomes magnetic under pressure (though the ordering is not known) but it also becomes metallic! It is a phase transition between a paramagnetic Kondo insulator to a magnetic metal.} 

\section{Model for the heterostructure}

\subsection{Essential requirements of parent compounds
% for realization of topological Kondo metals
 }
 We consider a heterostructure made of periodic stackings of SmB\textsubscript{6} and another, possibly magnetic isostructural compound $A$B\textsubscript{6} [see Fig.~\ref{fig1}~a)], candidates of which are discussed in Sec.~\ref{eq: candidates}.  
We will model parent compounds neglecting microscopic details as discussed in the next section. We consider one parent compound to be SmB\textsubscript{6}, because it is the most established candidate for a topological Kondo insulator. While the topological nature of SmB\textsubscript{6} has been under debate~\cite{stern2016}, our work is based on the assumption that SmB\textsubscript{6} is a topological Kondo insulator. Our results also transfer to other topological Kondo insulators.

The realization of topological Kondo semimetals in such a heterostructure depends on the presence of the requisite symmetries and the topological insulator electronic structure of SmB\textsubscript{6}, enabled by the Kondo effect.
As we will show, appropriate magnetic order then induces magnetic Weyl or Dirac cones in the electronic structure. These Weyl or Dirac cones may appear even if $A$B\textsubscript{6} and (or) SmB\textsubscript{6}---and with both (one) of them, the heterostructure---is metallic.~\cite{yan2017}. Despite metallicity not being ideal, we stress that all experimentally-confirmed Weyl semimetals thus far have in fact been metallic rather than semimetallic~\cite{yan2017}.

%\blue{(ref2: Recent experiments, including quantum oscillation measurements, have raised considerable doubts as to whether SmB6 is a topological Kondo insulator. That limits the practical feasibility of the proposed scheme.} 

%For the second parent compound $A$B\textsubscript{6}, we consider a hypothetical, topologically-trivial Kondo insulator. This allows us to carefully consider myriad rapid developments and uncertainty in understanding of the electronic structure of these candidates without restricting our calculations to what may turn out to be irrelevant scenarios at the level of microscopic details. The topological semimetal phases should be realized so long as the requisite symmetries and parent compound topologies protected by strongly-correlated Kondo physics are present in the heterostructure, even if the candidate parent compounds become metallic although this is not ideal~\cite{yan2017}.

\subsection{Treatment of correlations} 
\label{Sec: Treatment of correlations}
The topological properties of SmB\textsubscript{6} and \textit{A}B\textsubscript{6} important for realization of topological Kondo semimetals can be realized without taking into account the full multiplet structure of the $d$- or $f$- orbitals. We thus consider a simplified model that ignores these details. As the essential difference between topological Kondo phases and weakly-correlated topological phases, however, is that the former are protected by bulk electronic structure resulting from the Kondo effect as opposed to weakly-correlated band structure~\cite{chen2016}, we must construct topological Kondo semimetal phases using a model for parent topological Kondo insulator compounds in which non-trivial topology emerges at \textit{finite} correlation strength.

We therefore consider a lattice model for a topological Kondo insulator with cubic symmetry~\cite{legner2014} for the description of the parent compounds SmB\textsubscript{6} and \textit{A}B\textsubscript{6}:
\begin{equation}
\begin{split}
H &=\sum_{\gamma,s}
\Biggl[
\sum_{\bs{r}} \tilde{\varepsilon}^{}_{\gamma} c_{\gamma,\bs{r},s}^{\dagger} c^{}_{\gamma, \bs{r},s} 
-\sum_{j=1}^3 
\sum_{\langle \bs{r},\bs{r}' \rangle_j} \tilde{t}^{(j)}_{\gamma}
  c^{\dagger}_{\gamma, \bs{r},s} c^{}_{\gamma, \bs{r}',s} \Biggr] \\ 
&+
\sum_{\gamma,  s, s', i}
\sum_{\bs{r}}
\left( \mathrm{i}\tilde{V} c_{\gamma,\bs{r},s}^{\dagger} \sigma^{i}_{ss'} c^{}_{\bar{\gamma},\bs{r}+\bs{e}_i,s'} + \rm{H.c.}\right) \\ 
&+
\sum_{\bs{r}} U c^{\dagger}_{f, \bs{r}, \uparrow} c^{}_{f, \bs{r}, \uparrow} c^{\dagger}_{f, \bs{r}, \downarrow} c^{}_{f, \bs{r},\downarrow} \ .
 \end{split}
\label{eq: lattice hamiltoniann}
\end{equation}
Here, $c^{\dagger}_{\gamma, \bs{r},s}$ $\left(c^{}_{\gamma, \bs{r},s} \right)$ denotes a creation (annihilation) operator for an electron in orbital $\gamma\in\{d,f\}$ with spin $s\in\{\uparrow, \downarrow\}$ on site $\bs{r}$ of the cubic lattice, and $\langle \cdot \rangle_j$ for $j\in\{1,2,3\}$ denotes pairs of first (NN), second (NNN), and third neighbors (NNNN), respectively. The vectors $\bs{e}_i$ connect nearest neighbors in the $i \in \{x, y, z \}$ directions. In the middle line of Eq.~\eqref{eq: lattice hamiltoniann},
$\bar{\gamma}=f$ ($\bar{\gamma}=d$) when $\gamma=d$ ($\gamma=f$). The parameters $\tilde{t}^{(j)}_{\gamma}$ stand for $j$-th nearest neighbor hopping integrals, $\tilde{\varepsilon}_{\gamma}$ the onsite-energy of $\gamma$ band, and $\tilde{V}$ the hybridization between $f$ and $d$ orbitals. The form of the hybridization as a parity-odd hopping term is a consequence of the opposite inversion eigenvalues of the $d$ and $f$ orbitals. Terms containing $U$ govern interactions and reflect the assumption that $f$ electrons locally interact via a Hubbard repulsion while the $d$ electrons are non-interacting.

We consider the case where interactions strongly renormalize band parameters but low-energy excitations are described by well-defined Fermi-liquid quasiparticles. We can take advantage of previous work by Legner et al.~\cite{legner2014} that used the Kotliar-Ruckenstein slave-boson scheme in the mean-field approximation~\cite{kotliar1986} to treat interactions in the quasiparticle approximation to the periodic Anderson model~\cite{read1983, coleman1984, rice1985, kotliar1986}, assuming a $\bs{k}$-independent self-energy for the $f$-electrons of the Fermi liquid type $\Sigma_f \left( \omega \right) = a + b \omega + \mathcal{O} \left( \omega^2 \right)$. The Fermi-liquid quasiparticles in such a state are then accurately described by a non-interacting Hamiltonian with renormalized parameters $\tilde{t}^{(j)}_{f} \rightarrow t^{(j)}_f=z^2 \tilde{t}^{(j)}_{f}$, $\tilde{V} \rightarrow V = z\tilde{V}$, $\tilde{\varepsilon}_f \rightarrow \varepsilon_f=\tilde{\varepsilon}_f+\lambda$, and $U \rightarrow 0$, while the others remain the same. The additional parameters $z$ and $\lambda$ are expressed by the self-energy expansion coefficients  by $z=(1-b)^{-1/2}$ and $\lambda=a/(1-b)$, thus depend on the band parameters of the original Hamiltonian containing the quartic interaction terms between $f$-orbital electrons. The predominant effect of $U$ is to move the $f$-electron band closer to the Fermi level, thereby enabling topological band inversion.

\subsection{Low-energy effective theory for the heterostructure} 

%\blue{(ref2: Kondo/Anderson lattice models give rise to strong renormalizations as a result of the large Coulomb repulsion. Such effect does not seem to have been analyzed in the paper.}

Having discussed the effective non-interacting Hamiltonian that we use to describe bulk SmB\textsubscript{6} and $A$B\textsubscript{6}, we now study the mechanism by which three-dimensional (3D) bulk Weyl and Dirac cones arise in the heterostructure by weakly coupling the topological surface states of SmB\textsubscript{6} between consecutive interfaces within an effective theory for the surface states as expected in a heterostructure as shown in Fig.~\ref{fig1}(a).
Let us first consider both SmB\textsubscript{6} and the trivial insulator $A$B\textsubscript{6} to be nonmagnetic.
In the surface Brillouin zone (BZ) of SmB\textsubscript{6}, there are three Dirac cones appearing at the two \barpt{X} points  $(\pi,0),(0,\pi)$ and at the \barpt{\Gamma} point  $(0,0)$  [see Fig.~\ref{fig1} b)].
In the limit of decoupled interfaces, each interface between the two materials then hosts three Dirac cones, deriving from those appearing in the surface BZ of SmB\textsubscript{6}.

Weakly hybridizing these Dirac cones at $\Omega \in\{\barptt{X},\barptt{\Gamma}\}$ with overlap integrals $\Delta_\Omega$ and $\delta_\Omega$ across the SmB\textsubscript{6} layer and the $A$B\textsubscript{6} layer, respectively, yields a low-energy effective theory for the heterostructure of the form 
\begin{equation}
\begin{split}
\mathcal{H}_{\Omega}(\bs{k})
=& \,
v_{\mathrm{F}} \tau^z \left( \hat{\bs{z}} \times  \bs{\sigma} \right) \cdot (\bs{k}-\bs{k}^{\Omega}) 
+ \Delta_\Omega \tau^x \sigma^0
\\
&+ \delta_\Omega \left(\tau^x \sigma^0 \cos\,k_z
+ \tau^y \sigma^0 \sin\,k_z\right)
\end{split}
\end{equation}
with $\bs{k}^{\barptt{X}}=(\pi,\pi,0)$ and $\bs{k}^{\barptt{\Gamma}}=(0,0,0)$.
Here, $\bs{\sigma}$ and $\bs{\tau}$ denote the vector of Pauli matrices in spin space and in the space of interfaces $\eta\in\{\text{upper}, \text{lower}\}$, respectively. 
Similarly, $\sigma^0$ and $\tau^0$ are the $2 \times 2$ unit matrices in spin space and in the space of interfaces, respectively.
For simplicity, we assume all three Dirac cones to be at the same energy and to be isotropic with the same Fermi velocity $v_\mathrm{F}$~\cite{roy2015}.

We first discuss the low-energy theory without magnetic order.
As translational symmetry is not broken in the $xy$ plane, the findings of \Ref{burkov2011} apply separately for each of the Dirac cones at the surface of SmB\textsubscript{6}.
Therefore, a band inversion occurs on the $\Gamma$--{Z} line when $|\Delta_{\barptt{\Gamma}}|<|\delta_{\barptt{\Gamma}}|$, and on the X--R line when $|\Delta_{\barptt{X}}|<|\delta_{\barptt{X}}|$.
As we tune the number of SmB\textsubscript{6} layers ($N_{\textrm{TI}}$) and $A$B\textsubscript{6} layers ($N_{\textrm{BI}}$), the ratios $|\Delta_{\barptt{\Gamma}}/\delta_{\barptt{\Gamma}}|$ and $|\Delta_{\barptt{X}}/\delta_{\barptt{X}}|$ change, thus making it possible to observe multiple topologically distinct phases. Depending on whether or not band inversions occur on the two different high-symmetry lines, we can distinguish four different phases [see Fig.~\ref{fig:fig2} a)]: A weak topological insulator (WTI) if there is a band inversion along the X--R line, a strong topological insulator (STI\textsubscript{2}) for a band inversion along the $\Gamma$--Z line, another strong topological insulator (STI\textsubscript{1}) for band inversions on both lines, or a band insulator (BI) without any band inversions.
Considering the extreme limits of either only SmB\textsubscript{6} or only $A$B\textsubscript{6}, we expect the system to be in the phases STI\textsubscript{1} and BI, respectively. Therefore, it is expected that the system will realize either of two phase transitions STI\textsubscript{1}--WTI--BI or STI\textsubscript{1}--STI\textsubscript{2}--BI as we increase $N_{\textrm{BI}}/N_{\textrm{TI}}$ from $0$ to $\infty$ [see Fig.~\ref{fig:fig2} a)].

We now consider the heterostructure with magnetic order.
In the case of ferromagnetic order in the $\hat{z}$ direction, there is no coupling between Dirac cones centered around different points in the BZ, so we may consider each one individually with an additional term for the net magnetization in the $\hat{z}$ direction:
\begin{equation}
\mathcal{H}^\text{FM}_{\Omega}(\bs{k}) = \mathcal{H}_{\Omega}(\bs{k}) +  m \tau^0 \sigma^z.
\label{FM Hamiltonian}
\end{equation}
The eigenvalues of this Hamiltonian are given by
\begin{subequations}
\begin{align}
E_{\alpha, \beta}^{\Omega} &= \alpha \sqrt{v_{\textrm{F}}^2 (\tilde{k}_x^2+\Delta \tilde{k}_y^2)+ [ m + \beta M_{\Omega}(k_z) ]^2} ,
\end{align}%
with
\begin{align}
\tilde{k}_x= k_x-k_x^{\Omega}, \quad \tilde{k}_y=k_y-k_y^{\Omega},
\end{align}
\begin{align}
M_{\Omega}(k_z) &= \sqrt{\Delta_\Omega^2 + \delta_\Omega^2 +2 \delta_\Omega \Delta_\Omega \cos \, k_z} ,
\end{align}%
\label{low-energy eigenvalues}%
\end{subequations}%
where $\alpha$ and $\beta$ can take the values $\pm1$.
Thus, at $k_x=k_x^{\Omega}$, $k_y=k_y^{\Omega}$, $m=\pm M_{\Omega}(k_z)$, we obtain Weyl cones.

As a second case, we discuss antiferromagnetic order with an ordering vector $\bs Q=(\pi,\pi,0)$. (Note that $Q_z$ is defined with respect to the unit cell of the heterostructure, not the lattice spacing of the comprising materials.)
In that case, the surface Dirac cones at the two \barpt{X} points are coupled, leading to the Hamiltonian
\begin{equation}
\mathcal{H}^\mathrm{{AFM}}_{\barptt{X}}(\bs{k}) =
\rho^0 \mathcal{H}_{\barptt{X}}(\bs{k}) + m \rho^x \tau^0 \sigma^z,
\label{eq: AFM hamiltonian}
\end{equation}
where $\bs{\rho}$ and $\rho^0$ are the vector of Pauli matrices and the $2 \times 2$ unit matrix in the space of the two Dirac cones at the different $\barptt{X}$ points, respectively.
This Hamiltonian has the same eigenvalues as those given in Eq.~\eqref{low-energy eigenvalues} with $\Omega=\barptt{X}$, but with an extra double degeneracy of each band.

The antiferromagnetic order breaks time-reversal symmetry $T$, but preserves the combination of $T$ and a translation $t_{\bs{a}}$ by a displacement $\bs{a}=(1,0,0)$ in units of the lattice constant. The symmetry $(T t_{\bs{a}})$ is represented by $-\mathcal{K} \rho^z \tau^0 \sigma^y$ (where $\mathcal{K}$ represents complex conjugation) and in combination with inversion symmetry $I$ leads to Kramer's degeneracy at each $\bs{k}$. In addition, the heterostructure has a $C_4^z$ rotation symmetry that protects Dirac cones on the $C_4^z$ invariant lines, such as $k_x=k_y=0$, as a crossing of two pairs of bands with different $C_4^z$ eigenvalues.
% $(t_{\bs{a}} T)\mathcal{H}_{\mathrm{AFM}}(\bs{k})(t_{\bs{a}} T)^{-1}=\mathcal{H}_{\mathrm{AFM}}(-\bs{k})$. It furthermore preserves the $C_4^z$ rotation symmetry and inversion symmetry $I$ of the heterostructure. 
%For Hamiltonian~\eqref{eq: AFM hamiltonian}, these symmetries are represented by
%\begin{equation}
%\begin{split}
%T=&\mathcal{K} \rho^0 \otimes \tau^0 \otimes \sigma^y \\
%t_a=&-\rho^z \otimes \tau^0 \otimes \sigma^0 \\
%C_4^z=& -\rho^x \otimes \tau^0 \otimes e^{i \frac{\pi}{4} \sigma^z} \\
%I=& \rho^0 \otimes \tau^x \otimes \sigma^0
%\end{split}
%\end{equation}
%The product of inversion $I=\sigma^{0} \otimes \sigma^{x} \otimes \sigma^{x} \otimes \sigma^{0}$, translation $t=\sigma^{z} \otimes \sigma^{0} \otimes \sigma^{0} \otimes \sigma^{0}$, and time-reversal $T= \mathcal{K} \sigma^{0} \otimes \sigma^{0} \otimes \sigma^{0} \otimes \sigma^{y}$ gives us doubly-degenerate band structure. In addition, the 4-fold rotation $C_{4}=\sigma^{x} \otimes \sigma^{0} \otimes \sigma^{0} \otimes (e^{i\pi/4 Ksigma^{z}})$ protects the bulk Dirac cones in the heterostructure.
For finite $(\pi,\pi,0)$ magnetic order and $\Delta / \delta \sim\pm 1$ , bulk Dirac cones emerge in pairs near the A and M point in the folded  BZ, respectively.

\subsection{Lattice model}

%\blue{(ref2: Kondo/Anderson lattice models give rise to strong renormalizations as a result of the large Coulomb repulsion. Such effect does not seem to have been analyzed in the paper.}

The effective model for the heterostructure showed that Weyl and Dirac cones can appear in the ferro- and antiferromagnetically ordered case, respectively, but it cannot be used to determine the strength of the hybridization terms $\Delta_\Omega$ and $\delta_{\Omega}$. To determine which of the possible phases can be realized in the actual heterostructure, we now use the quasiparticle Hamiltonian for a topological Kondo insulator with cubic symmetry, which contains parameters renormalized by the Hubbard $U$~\cite{legner2014}:
\begin{multline}
% \begin{split}
H =\sum_{\gamma,s}
\Biggl[
\sum_{\bs{r}} \varepsilon^{}_{\gamma,\bs{r}} c_{\gamma,\bs{r},s}^{\dagger} c^{}_{\gamma, \bs{r},s} 
-\sum_{j=1}^3 
\sum_{\langle \bs{r},\bs{r}' \rangle_j} t^{(j)}_{\gamma}
  c^{\dagger}_{\gamma, \bs{r},s} c^{}_{\gamma, \bs{r}',s} \Biggr] \\
+
\sum_{\gamma,  s, s', i}
\sum_{\bs{r}}
\left( \mathrm{i}V c_{\gamma,\bs{r},s}^{\dagger} \sigma^{i}_{ss'} c^{}_{\bar{\gamma},\bs{r}+\bs{e}_i,s'} + \rm{H.c.}\right)  \ .
% \end{split}
\label{eq: lattice hamiltonian}
\end{multline}

To model the heterostructure, we endow the onsite energies $\varepsilon^{}_{\gamma,\bs{r}} $ with a spatial dependence to change their values between the two materials comprising the heterostructure. The parameter $\varepsilon^{}_{f}-\varepsilon^{}_{d} $ changes the bulk band topology of the translationally invariant model between normal insulator, weak topological insulator, and strong topological insulator~\cite{legner2014,dzero2016}.
As discussed in Sec.~\ref{Sec: Treatment of correlations}, the effect of correlations is encoded in the renormalization of the model parameters: most notably, increasing the correlation strength induces a shift in $\varepsilon^{}_{f}$ and reduces $V$, $t^{(1)}_f$, and $t^{(2)}_f$~\cite{legner2014}. 
 
The heterostructure is then modeled as $N_{\rm{TI}}$ layers of the topological Kondo insulator SmB\textsubscript{6}, using the parameter set~\cite{legner2014}  $t^{(1)}_f=-0.1 t^{(1)}_d$, $t^{(2)}_d=-0.4 t^{(1)}_d$, $t^{(2)}_f=0.04 t^{(1)}_d$, $t^{(3)}_d=t^{(3)}_f=0$, $V=0.5 t^{(1)}_d$, $\epsilon_d=1.76 t^{(1)}_d$, $\epsilon_f=-0.24 t^{(1)}_d$, where we set $t^{(1)}_d=\SI{100}{\milli\eV}$ in order to produce the correct band gaps,
and $N_{\rm{BI}}$ layers of a trivial Kondo insulator with the same parameters except for $\epsilon_d=4.02 t^{(1)}_d$, $\epsilon_f=-0.78 t^{(1)}_d$, corresponding to our example $A$B\textsubscript{6}. The bulk band structures of these two parameter sets are shown in Fig.~\ref{fig1}~c) and~d). The (topological) band gap of SmB\textsubscript{6} is then \SI{24}{\meV}, a value in good agreement with past work~\cite{chang2015}. For $A$B\textsubscript{6}, we choose a  band gap of \SI{84}{\meV}.

\section{Results}

\subsection{Phase diagram at zero magnetization}
We first explore the topology of the heterostructure in the absence of magnetic order.
We fix the number of layers of SmB\textsubscript{6}, $N_\mathrm{TI}$ ($N_\mathrm{TI}=2$ in our examples), and characterize the topology of the heterostructure as a function of the number of layers of $A$B\textsubscript{6}, $N_\mathrm{BI}$. We can therefore interpolate between the limits of the topological Kondo insulator SmB\textsubscript{6} for $N_\mathrm{BI}=0$ and YbB\textsubscript{6} for $N_\mathrm{BI}\gg N_\mathrm{TI}$.
The chemical potential is fixed to lie between the two middle bands of the heterostructure dispersion, corresponding to half filling.

\begin{figure}[t]
\centering
\includegraphics[width=0.47\textwidth]{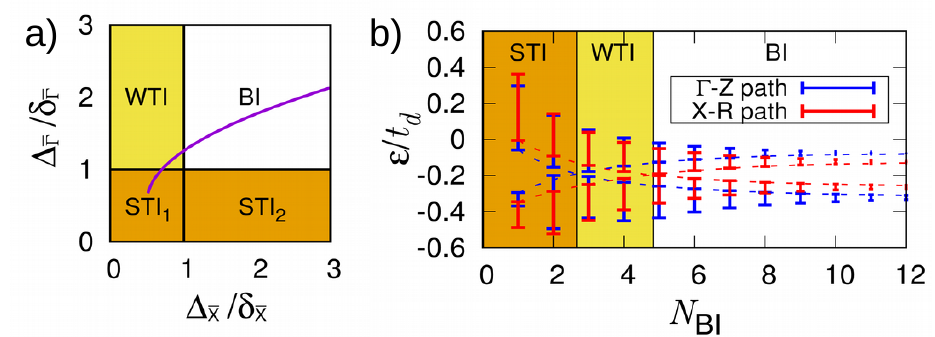}
\caption{
Topological phases of the nonmagnetic heterostructure.
a)
Phase diagram of the effective model for the heterostructure based on
the coupling of topological Dirac surface states. Depending on the strength of the couplings between the 
surface states, $\Delta_{\barptt{\Gamma},\barptt{X}}$ and $\delta_{\barptt{\Gamma},\barptt{X}}$, the heterostructure can realize a normal band insulator (BI), a weak topological insulator (WTI), or a strong topological insulator with (STI\textsubscript{1}) or without (STI\textsubscript{2}) nontrivial weak indices.
b)
Phase diagram of the tight-binding model for the heterostructure with a fixed number of 2 layers of SmB\textsubscript{6} and a variable number $N_{\mathrm{BI}}$ of layers of the trivial Kondo insulator $A$B\textsubscript{6}. Vertical bars indicate the band widths of the highest valence and lowest conduction band along the high-symmetry lines $\Gamma$--Z and X--R. Topological transitions are accompanied by gap closings at the end of these lines, indicated by the dashed guides to the eye. The system follows a trajectory similar to the blue line in the phase diagram a) as $N_{\mathrm{BI}}$ is varied.  
}
\label{fig:fig2}

\end{figure}

To show when band touchings occur, we compute where the minima and maxima of the middle two bands on high symmetry lines $\Gamma$--{Z} and X--R are located in energy as a function of the number of layers of $A$B\textsubscript{6}. Furthermore,  we compute the strong and weak  $\mathbb{Z}_2$ topological indices to determine the topological phases between band touchings. A series of phase transitions STI\textsubscript{1}--WTI--BI is detected between $N_{\textrm{BI}}=2$ and $3$, and between $4$ and $5$, respectively [see Fig. \ref{fig:fig2} b)]. We first focus on the former case and construct a phase diagram showing the number of Weyl cones formed between the two middle bands as a function of the number of $A$B\textsubscript{6} layers and the magnitude of the magnetization [see Fig.~\ref{fig3} a)].

\subsection{Ferromagnetism and the magnetic Weyl Kondo semimetal}
Ferromagnetic order could emerge via various mechanisms. In this work, we consider two cases: ferromagnetic order in SmB\textsubscript{6} or in the trivial Kondo insulator (corresponding to, e.g., ferromagnetic EuB\textsubscript{6}).
\begin{figure}[t]
\centering
\includegraphics[width=0.46\textwidth]{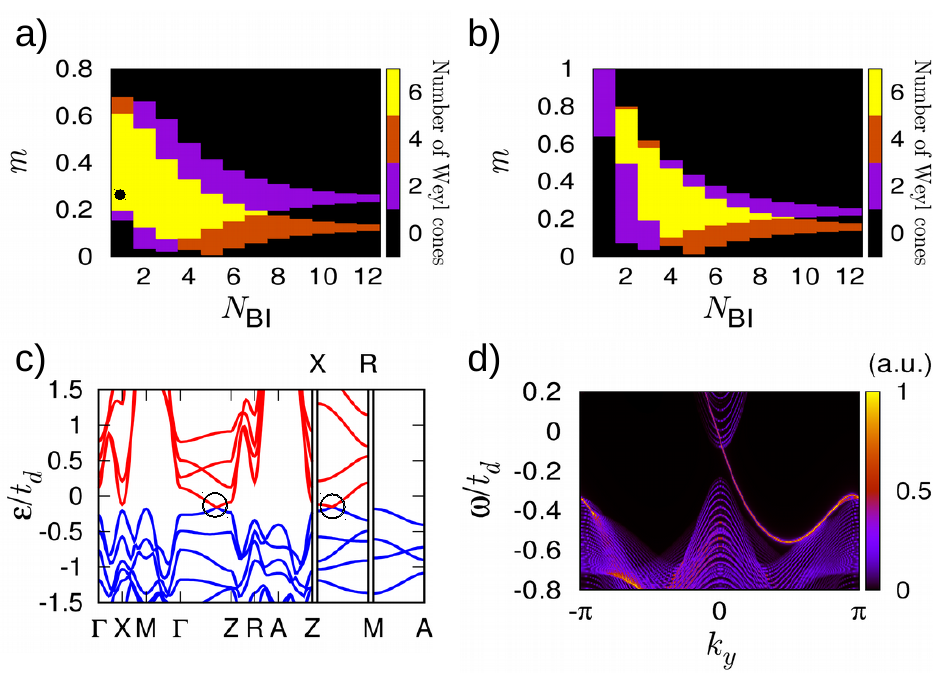}
\caption{
Weyl semimetal phases of the ferromagnetic heterostructure.
a), b) Phase diagrams indicating the number of Weyl cones between the highest valence and the lowest conduction band for ferromagnetic order a) in the SmB\textsubscript{6} and b) in the $A$B\textsubscript{6} layers. Here, $m$ is the strength of the magnetization and the number of SmB\textsubscript{6} layers is fixed to 2. 
c) Band structure of the heterostructure for parameter values corresponding to the black circle in a).
Weyl cones are indicated by black circles.
d)
Surface spectral function of the (100) surface for a slab geometry of the heterostructure in c) at $k_z=\pi$, showing a chiral mode belonging to a Fermi arc surface state.
}
\label{fig3}

\end{figure}

We find four different phases depending on $N_{\textrm{BI}}$ and $m$, corresponding to $0$, $2$, $4$, and $6$ Weyl cones present in the system, which are located along the high-symmetry lines $\Gamma$--Z and X--R. 
Depending on the parameters, these Weyl cones can either be of type I or type II~\cite{soluyanov2015,xu2015''}.
A representative band structure for magnetization $m=0.25t_d^{(1)}$ is shown in Fig.~\ref{fig3}~c). We can understand the phase diagram structure as follows: The Dirac cones of the $(001)$ surface of SmB\textsubscript{6} occur at the $\barptt{X}$ and $\barptt{\Gamma}$ points in the surface BZ and each Dirac cone can yield a pair of Weyl cones for finite magnetization in the layering direction, as discussed in the section on the low-energy effective theory of the heterostructure. The offset in energy between the surface Dirac cones of SmB\textsubscript{6} means six Weyl cones emerge at intermediate magnetization from the $\bar{\Gamma}$ and also each $\bar{\textrm{X}}$ surface Dirac cone. As $N_{\textrm{BI}}$ increases, the band width of the middle bands decreases, finally eliminating this intermediate regime and generating two tails where Weyl cones appear either only around the X points ($4$ Weyl cones) or the $\Gamma$ point ($2$ Weyl cones).

We also compute the spectral function of the heterostructure in a slab geometry, using again the magnetization $m=0.25t_d^{(1)}$.
With the heterostructure stacked in the $\hat{z}$ direction, we open the system in the $\hat{x}$ direction and compute the surface spectral function $A_{\textrm{surf}}(\omega,k_y,k_z) = \Im \Tr [\mathcal{G}(\omega,k_y,k_z) P_{\textrm{surf}}]$, where $\mathcal{G}(\omega,k_y,k_z)$ is the Green's function and $P_{\textrm{surf}}$ the projector on one surface layer.
This result is presented in Fig.~\ref{fig3}~d), where a Fermi arc, a signature of a Weyl semimetal~\cite{xu2015}, is clearly visible. 

We can similarly compute the dispersion, phase diagram, and spectral function for the case of a net magnetization in the trivial Kondo insulator, $A=\text{Eu}$.
The same three topologically non-trivial phases---characterized by $2$, $4$, and $6$ Weyl cones, respectively---that appear in Fig.~\ref{fig3}~a) also occur in this second phase diagram, shown in Fig.~\ref{fig3}~b).
%They occur over regions in phase space that are comparable in size, indicating the phases are largely unaffected by whether magnetization is derived from the trivial or topological Kondo insulator layers. However, topologically non-trivial regions of the phase diagram~\ref{fig3} b)  are slightly smaller than those of~\ref{fig3} a) , which may be due to the fact that the the effective magnetization strength seen by the topologically non-trivial surface states of SmB\textsubscript{6} is less than the actual magnetization strength of the trivial Kondo insulator layers.

\subsection{Antiferromagnetism and the magnetic Dirac Kondo semi\-metal}
Motivated by the low-energy model of the heterostructure for antiferromagnetic order, we now consider the full heterostructure Hamiltonian with finite magnetization in SmB\textsubscript{6} layers corresponding to antiferromagnetic order oriented in the stacking direction. 
We physically motivate this case as follows: There is evidence of antiferromagnetic order induced in bulk SmB\textsubscript{6} by compression~\cite{Barla2005, Derr2008}. Also, Eu$_{x}$Ca$_{1-x}$B\textsubscript{6}, where $x=0.4$ and $0.6$, was reported to show intrinsic antiferromagnetism below $\SI{3}{\K}$~\cite{rhyee2004}.

For the antiferromagnetic order, we add
\begin{subequations}
\begin{align}
\sum_{\gamma,s}
\sum_{\bs{r} \in \textrm{TI/BI}}
m^{\textrm{AFM}}_{\gamma,\bs{r},s} c_{\gamma,\bs{r},s}^{\dagger} c^{}_{\gamma, \bs{r},s}
\end{align}
to the Hamiltonian~\eqref{eq: lattice hamiltonian}, where $\bs{r}$ sums over either TI or BI layers depending on the case, and $m^{\textrm{AFM}}_{f,\bs{r},s}=s(-1)^{i+j+k}m'$ at $\bs{r}=i\bs{e}_{x}+j\bs{e}_{y}+k\bs{e}_{z}$.
We take a particular form of $m^{\textrm{AFM}}_{d,\bs{r},s}=0.2 m^{\textrm{AFM}}_{f,\bs{r},s}$ if $\bs{r}$ is in TI layers, and $m^{\textrm{AFM}}_{d,\bs{r},s}=m^{\textrm{AFM}}_{f,\bs{r},s}$ if $\bs{r}$ is in BI layers, to express compression-based antiferromagnetism in topological Kondo insulator layers dominant in the $f$ orbitals, as well as exciton-based antiferromagnetism that is neither predominantly of $d$ or $f$ orbital character~\cite{murakami2002,zhitomirsky1999}, respectively.
We also add an NNN hybridization
\begin{align}
\sum_{\gamma,  s, s', i}
\sum_{\bs{r}}
\left[ \mathrm{i}V' c_{\gamma,\bs{r},s}^{\dagger} (\bs{\sigma} \cdot \hat{\bs{d}}_i)_{ss'} c^{}_{\bar{\gamma},\bs{r}+\bs{d}_{i,s'}} + \rm{H.c.}\right]
\end{align}
\end{subequations}
in this section to isolate Dirac cones in the heterostructure bulk, where $V'$ denotes the NNN hybridization coefficient, $\bs{d}_i$, $i=1,\dots ,6$, are taken from the set of six directed connections to the NNN sites, $ \{ \bs{e}_x \pm \bs{e}_y, \bs{e}_y \pm \bs{e}_z, \bs{e}_z \pm \bs{e}_x \} $, and $\hat{\bs{d}}_i$ denote the corresponding unit vectors. As representative values, we set $t^{(3)}_{\gamma}=0.16t^{(1)}_{\gamma}$ and $V'=-0.2t^{(1)}_d$.

\begin{figure}[t]
\centering
\includegraphics[width=0.46\textwidth]{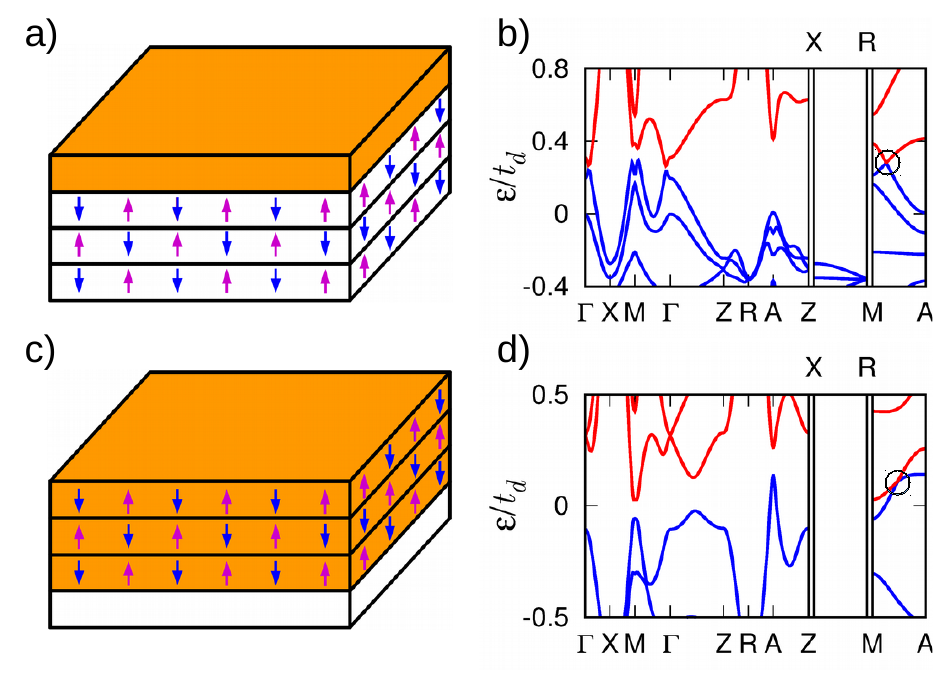}
\caption{
Antiferromagnetic configurations for cases a) $N_{\textrm{TI}}=1$, $N_{\textrm{BI}}=3$, $m'=1.0$ in BI layers, and c) $N_{\textrm{TI}}=3$, $N_{\textrm{BI}}=1$, $m'=0.5$ in TI layers, and corresponding dispersions for the full heterostructure tight-binding Hamiltonian are shown in b) and d), respectively. A type-I (type-II) Dirac cone is visible on the M--A high-symmetry line in the BZ in subfigure b (d), highlighted with a black circle.}
\label{fig4}
\end{figure}

Two examples of Dirac semimetal dispersions of the heterostructure are shown in Fig.~\ref{fig4}. Both type-I and type-II~\cite{chang2016} Dirac cones can occur in the full model, shown in Fig.~\ref{fig4} b) and d), respectively.

%\blue{ref1: Authors consider YbB6 as a trivial insulator. Even though there is a controversy for the nature of YbB6, there are some well established facts. YbB6 has fully filled f shell (f14) and as a result it has no f bands near the Fermi energy therefore the tight binding density of states in Fig 1(d) is wrong. Some papers even argue that it is a weakly interacting topological insulator due to band inversion between p and d bands.} 

\section{ $A$B\textsubscript{6} material candidates}
\label{eq: candidates}
For the Weyl Kondo semimetal case, a promising candidate for $A$B\textsubscript{6} is EuB\textsubscript{6} as it displays the requisite ferromagnetic order~\cite{degiorgi1997, sullow1998}. We note that there is evidence EuB\textsubscript{6} is metallic~\cite{hirsch1999} as well as studies suggesting the compound is a semimetal or a half-metallic semimetal~\cite{kune2004}, but emphasize again that this is not a problem for realization of the magnetic Weyl Kondo semimetal phase in practice: so long as the symmetries and parent compound topologies required for the magnetic Weyl Kondo semimetal phase are present, bulk Weyl cones will occur in the heterostructure near the Fermi level. We furthermore emphasize that Weyl cones in bulk dispersions have so far only been observed in what are actually metallic systems~\cite{yan2017}. We also note evidence that doping with Ca can push EuB\textsubscript{6} into an insulating state~\cite{urbano2005}.

For the magnetic Dirac Kondo semimetal case, we consider two promising candidates for $A$B\textsubscript{6}. First, we discuss the potential of YbB\textsubscript{6} for this purpose. We note that the potential of YbB\textsubscript{6} as a topological Kondo insulator has been studied: the 4$f$ and 5$d$ bands of YbB\textsubscript{6} were proposed to be closest in energy to the Fermi level and the compound was proposed as a topological Kondo insulator candidate~\cite{weng2014}. Subsequently, it was suggested that the B 2$p$ and Yb 5$d$ bands of YbB\textsubscript{6} were actually closest in energy to the Fermi level and the material was proposed to be a topological insulator candidate due to inversion between predominantly $p$ and $d$ orbital character bands~\cite{neupane2015}. This agreed with other work~\cite{kakizaki1993, xia2014, frantzeskakis2014} showing that the binding energy of the Yb 4$f$ $7$/$2$ band is about 1 eV below the Fermi level. More recently, evidence was found, which suggests YbB\textsubscript{6} has bands with predominantly $p$- and $d$-orbital character near the Fermi level but that it is a \textit{trivial insulator}~\cite{kang2016}, although the same study indicates the compound may become a $p$-$d$ overlap semimetal under pressure. We therefore note that YbB\textsubscript{6} may satisfy the topology and symmetry requirements necessary to serve as $A$B\textsubscript{6} for construction of the magnetic Dirac Kondo semimetal. Namely, it is topologically trivial~\cite{kang2016}, and we propose the magnetic structure of SmB\textsubscript{6} under pressure realizes the requisite symmetries for the magnetic Dirac Kondo semimetal phase.

CeB\textsubscript{6} is another promising candidate material: it exhibits a low-temperature magnetic phase~\cite{deursen1982, harrison1998, teklu2000, hall2000, zaharko2003, goodrich2004, plakhty2005, friemel2012, jang2014}, with antiferromagnetic order below 2.3~K~\cite{zaharko2003}. Angle-resolved photoemission data~\cite{neupane20152} furthermore revealed the presence of 4$f$ flat bands and dispersive 5$d$ bands in the vicinity of the Fermi level. Transport studies indicate CeB\textsubscript{6} behaves as a Kondo metal~\cite{neupane20152}, but we reiterate that the magnetic Dirac Kondo semimetal phase discussed here persists even if parent compounds become metallic, so long as the requisite symmetries and parent compound topologies hold, although this is not ideal~\cite{yan2017}.

\section{Discussion} We consider topological Kondo insulator heterostructures as platforms for the realization of magnetic Weyl and Dirac Kondo semimetal phases. We find ferromagnetism (antiferromagnetism) in the heterostructure generically realizes topologically-protected type-I and type-II Weyl (Dirac) cones sufficiently isolated from other states and proximate in energy to the Fermi level to realize type-I and type-II magnetic Weyl (Dirac) Kondo semimetal phases near half filling.

We note that thin films of SmB\textsubscript{6} have already been grown via molecular beam epitaxy (MBE)~\cite{shishido2015} and Kondo superlattices of other compounds have been grown via MBE for study of quantum criticality~\cite{ishii2016}. Furthermore, evaporation of boron and most rare-earth lanthanides is possible at operating temperatures for effusion cells~\cite{hoffman2016}. Negative pressure on SmB\textsubscript{6} (lattice constant $a = \SI{4.13}{\angstrom}$~\cite{jiang2016, roessler2014}) due to interfaces with EuB\textsubscript{6} ($a = \SI{4.19}{\angstrom}$~\cite{kreissl2005}), YbB\textsubscript{6} ($a = \SI{4.18}{\angstrom}$~\cite{kang2016}), or CeB\textsubscript{6} ($a = \SI{4.14} {\angstrom}$~\cite{mackinnon2013}) may also permit the observation of the desired Kondo physics at much higher temperatures of up to \SI{240}{\K} as well as enhancement of one or both parent compound band gaps given observed effects of tensile strain in SmB\textsubscript{6}~\cite{stern2016}. 

%Other recent experimental work reports highly spin-polarized surface states, indicating the surface states of SmB\textsubscript{6} are topologically-protected~\cite{kong2017}.

\begin{acknowledgments} 
The authors gratefully acknowledge helpful discussions with Jason Hoffman, Manfred Sigrist, Ronny Thomale, Ian Affleck, Marcel Franz, Arun Paramekanti, Victor Galitski and Piers Coleman. This project was supported by the Swiss National Science Foundation (grant no. 200021\_169061). AMC would also like to thank the Aspen Center for Physics, which is supported by National Science Foundation grant PHY-1066293, for hosting during some stages of this work.
\end{acknowledgments}

\bibliography{kwsm.bib}
\end{document}